# Publication and collaboration anomalies in academic papers originating from a paper mill: evidence from a Russia-based paper mill

Anna Abalkina, Freie Universität Berlin

anna.abalkina@fu-berlin.de

**Abstract**

This study attempts to detect papers originating from the Russia-based paper mill "International publisher" LLC. A total of 1009 offers published during 2019-2021 on the 123mi.ru website were analysed. The study allowed us to identify at least 434 papers that are potentially linked to the paper mill including one preprint, a duplication paper and 15 republications of papers erroneously published in hijacked journals. Evidence of suspicious provenance from the paper mill is provided: matches in title, number of coauthorship slots, year of publication, country of the journal, country of a coauthorship slot and similarities of abstracts. These problematic papers are coauthored by scholars associated with at least 39 countries and submitted both to predatory and reputable journals. This study also demonstrates collaboration anomalies and the phenomenon of suspicious collaboration in questionable papers and examines the predictors of the Russia-based paper mill. The value of coauthorship slots offered by "International Publisher" LLC in 2019-2021 is estimated at $6.5 million. Since the study analysed a particular paper mill, it is likely that the number of papers with forged authorship is much higher.

**Keywords**: ghostwriting, paper mills, academic misconduct, coauthorship for sale, suspicious collaboration, hijacked journals

## Introduction. Manufactured misconduct

Paper mills represent an offer or on-demand writing of fraudulent academic manuscripts for sale. Paper mills also provide additional services, such as searches for coauthors, submission of manuscripts, revision and control of publications and indexation of the paper in international databases. Very often, fraudulent entities selling coauthorship slots or entire academic papers mimic companies offering text-editing services or translation services (Hvistendahl 2013). However, the cost of such paper mill production significantly exceeds the cost of real editing services.

The frequency rate of papers originating from paper mills in the academic literature is unknown. Recent investigations by research integrity experts have shown the infiltration of the academic literature with paper mill production (Bik 2020, Schneider 2020). Since 2021 journals initiated mass retractions. In January 2021, the Royal Society of Chemistry announced a series of retractions by its journals. RSC Advances retracted 68 papers due to the "systemic production of falsified research", and Food and Function and RSC Medicinal Chemistry retracted one paper each (RSC, 2021). All of these papers were submitted by authors at Chinese hospitals, had common structures and templates and were assumed to be productions of paper mills (Else & Van Noorden, 2021). In December 2021, SAGE retracted 122 papers because of submission or peer-review manipulations associated with paper mill production (Oransky 2021). In February 2022, IOP Publishing has retracted 350 papers at once from two conference proceedings due to the lack of peer-review, citation manipulations, presence of "tortured phrases" discovered by Cabanac et al. (2021), and text similarities (Oransky 2022).

According to the Retraction Watch database, since 2020, massive retractions of papers originating from paper mills have occurred (Figure 1). As of December 2021, 3450 fraudulent manufactured papers have been identified[1]. However, this discovery could be just the tip of the iceberg because paper mills act on an anonymous basis, and their production cannot be easily detected.

**Figure 1**

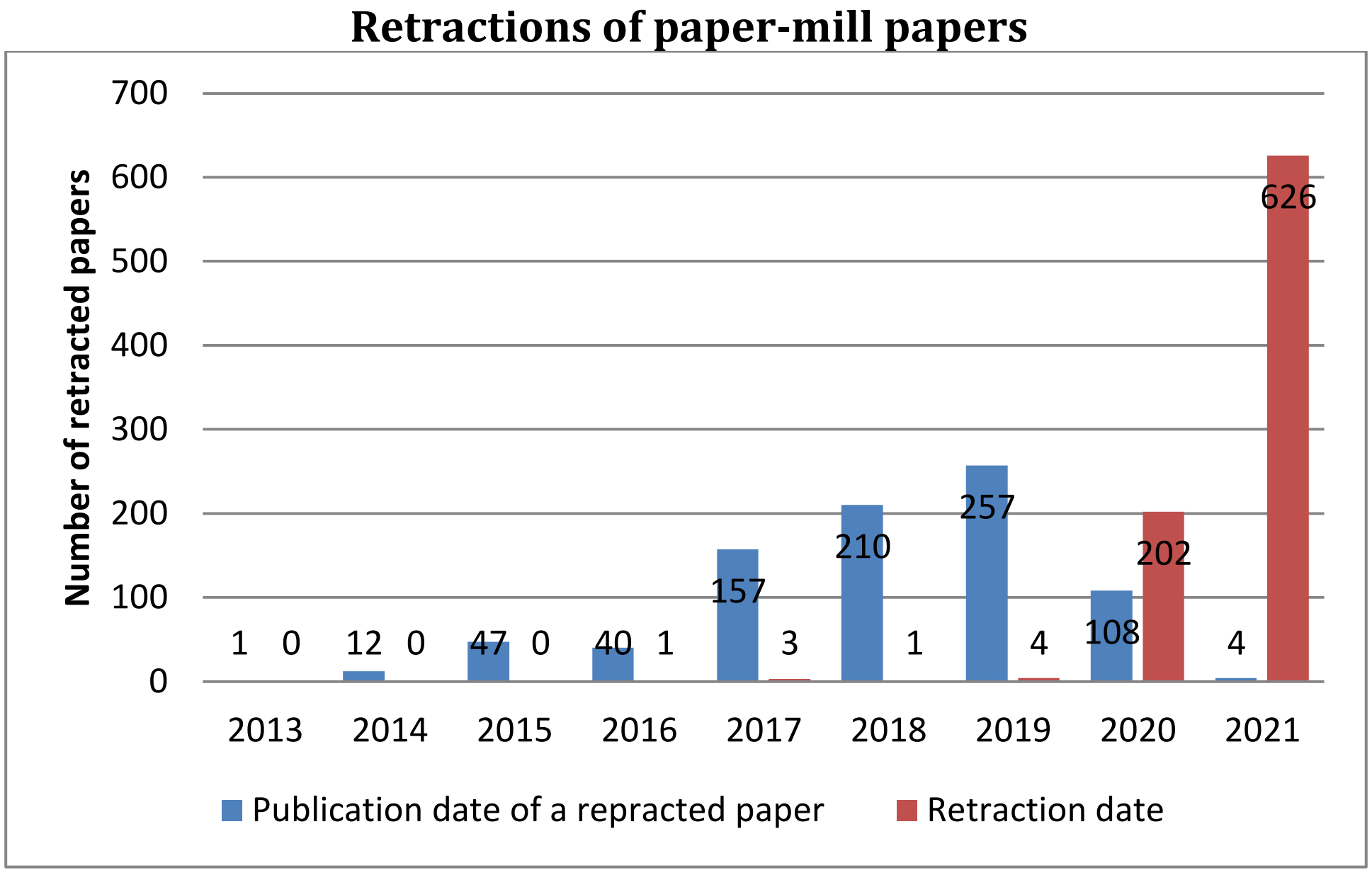

Source: Retraction Watch. URL: https://retractionwatch.com/retraction-watch-database-user-guide/ manipulations with peer review

To date, paper mills have been detected due to suspicious submission process or due to anomalies in the paper, falsification/fabrications in images and data, and similarities between texts. The manipulations during the submission process include falsification of peer-review processes (COPE Council 2021, Grove 2021, Day 2022), numerous submissions to a journal with the same patterns (geographical region, the same email linked to different accounts), request of the authorship changes after acceptance of the manuscript (COPE Council 2021).

[1] https://docs.google.com/spreadsheets/d/1zKxfaqug4ZhwHyGzslF38pFyC8xtU8lzmmOFMGYITDI/edit#gid=0

Streamlined production of dishonest papers is also associated with the usage of common templates despite not sharing the same coauthors, such as paper structure and similar section titles (Byrne & Christopher 2020, Cabanac et al. 2021, Else & Van Noorden 2021, Heck et al. 2021, RSC 2021), similar formatting (Byrne & Labbé 2017, Byrne & Christopher 2020), similar colours and types of diagrams (Cabanac et al. 2021), and identical fonts on figures (Byrne & Labbé 2017). Articles from paper mills could demonstrate other discrepancies that could potentially draw suspicion of non-authentic authorship. Recent evidence has also suggested suspicious authorship in papers of questionable provenance, such as rare individual authorship (Retraction Watch 2016), lack of previous publications on the topic of the paper (Retraction Watch 2016), unnatural collaborations between coauthors from different universities (RAS 2020), and suspicious affiliations, e.g., a university that unlikely supports certain types of experiments or research (Schneider 2021).

The production of paper mills demonstrates systematic violation of academic ethics (Christopher 2018), e.g., fabrication and falsification of data (Else & Van Noorden 2021), fabrication of images and western blots (Christopher 2018, van der Heyden 2021), plagiarism (Retraction Watch 2016, RAS 2020), and citation manipulations (Christopher 2021).

The majority of known paper mills originate from China (Hu & Wu 2013, Hvistendahl 2013, Liu & Chen 2018, Schneider 2020) and operate predominantly in medicine because of publication requirements for the promotion of practising doctors. There is evidence of paper mill operation in other countries, namely Iran and Russia (Stone 2016, Else & Van Noorden, 2021, Abalkina 2021). However, there is still little known about their activities: strategies to attract potential authors, submission strategies or even collusion with editor or journals. The goal of this study is to shed light on the activity of paper mills using the experience of the Russia-based one, to identify the fraudulent papers originating from them and to detect a set of predictors of fraudulent papers.

## Unethical publication practises in Russia

Paper mills offering coauthorship slots in papers submitted to international journals is a rather new phenomenon in Russia. They appeared as a response to the new regulatory framework of 2011-2012, setting new criteria for research evaluation, including publications and citations in international journals indexed in Web of Science and Scopus (Abalkina 2021) and setting nationwide indicators. According to President Putin's May decrees of 2012, the share of publications by Russian scholars among the total number of publications in scientific journals index in Web of Science should reach 2.44% by 2015, and at least five Russian universities should be ranked in the top 100 world's leading universities by 2020[2]. These legal acts shaped the academic landscape for the subsequent decade. First, the 5-100 project, also known as the Russian Academic Excellence Project, selected 21 Russian universities to enter international top rankings, which also meant increasing their publication performance. Second, in response to the new legislative framework, universities introduced new publication criteria for effective contracts, promotion or financial benefits.

Unfortunately, in addition to the positive effects of the new framework, there were some negative consequences of questionable or even fraudulent behaviour by scholars. First, a significant increase in publications in predatory journals was registered. RAS (2020) found numerous publications with plagiarism or questionable authorship in papers submitted to predatory journals by Russian scholars. Second, a significant increase in paper mills offering coauthorship for sale was discovered.

In Russia, there is a long-standing base for the illegal or unofficial marketing of papers written for submission in international journals. Recent work by Davydov and Abramov (2021) showed the blatant scale of contract cheating, by both students and faculty members. Evidence has also suggested that, over several decades, dozens of dissertation mills have flourished in Russia (Rostovtsev 2017, Abalkina 2020). More than 10,000 dissertations with massive plagiarism were detected by Dissernet, a grassroots initiative to address the issue of plagiarism in PhD theses and academic papers. Journals were well integrated into these dissertation mill schemes due to the requirements for publication of dissertation research results. The Russian Academy of Sciences Commission for Counteracting the Falsification of Scientific Research investigated 2528 papers with plagiarism and "obscure" coauthorship in 541 Russian journals. Journals retracted more than 800 papers after communication by the Commission (Chawla 2020). Later, a report by the Russian Academy of Sciences (RAS 2020) observed the reorientation of dishonest activity from producing plagiarized PhD theses to shadow marketing of academic papers.

## "International publisher" LLC

There are dozens of advertisements on the internet that approach Russian scholars to purchase coauthorship in a paper that will be submitted to a journal indexed in Scopus or

[2] https://rg.ru/2012/05/09/nauka-dok.html

Web of Science. It is difficult to confirm that all of these sales of coauthorship truly occur. These offers could be fraud to collect money from scholars and do not guarantee publication of the paper. Among these companies, “International Publisher” LLC offers coauthorship for sale and guarantees publication. However, is it a real paper mill company? A correspondent for the Russian online media The Insider documented a test purchase from “International Publisher” LLC of coauthorship in a paper, which was supposed to be published in 2019 (Litoy 2019). The paper, entitled “Project-Based Learning as a Tool for the Formation and Development of the Entrepreneurial Skills of Students” was indeed published in the Journal of Entrepreneurship Education*,* together with several coauthors. The paper was devoted to surveying students in Omsk, Russia, but none of the coauthors live or work in the Omsk region. Moreover, among all of the coauthors on the paper, none specializes in education studies (among the coauthors, there are journalists and scholars specializing in chemistry, history, and engineering). Some of the coauthors confirmed that they purchased a coauthorship slot at “International Publisher” LLC (Litoy, 2019).

“International Publisher” LLC is one of the best known companies in Russia that offers coauthorship for sale. It is a registered legal entity and has an office in one of the modern skyscrapers in the Moscow International Business Center (Moscow-City).

“International Publisher” LLC claims on its website that approximately 20,000 scholars published 4,000 papers in journals indexed in Scopus or Web of Science with the intermediation of the company. Offers to purchase coauthorship are openly listed on the 123mu.ru website (in Russian) or on the website with a rather clearer name: http://buy-sell-article.com/coauthorship.php#banner (multilingual version). As of mid-March 2022, the website listed 2376 papers with between one and five authorship slots to purchase.

In this paper auction, one can choose the topic of the paper, his or her position in the list of authors, the quartile of the journal, the date of publication, and the database where the journal will be indexed. As of mid-March 2022, 2320 papers for sale should have been indexed in Scopus, 480 papers in Web of Science, and 37 on Russian lists of journals compiled by the Higher Attestation Commission (VAK list).

The papers offered cover several disciplines, such as economics, law, education, linguistics, medicine, engineering, and agriculture. All of these areas (except linguistics) are considered to be the most corrupt in Russia. According to Dissernet’s data, economics, education, law, medicine, and engineering represent 83% of all detected plagiarized PhD theses in Russia (www.dissernet.org).

According to the website, an author should not worry about anything; "International Publisher" LLC will take care of the entire process of publication and indexation of the manuscript with the name of the client. S/he needs only pay. The price range for coauthorship varies from 180 euros to 5,000 euros. The price depends on the position of the coauthor (1st coauthorship costs the most) and the impact factor and reputation of the journal. For example, according to the offer, the highest price (5,000 euros) is charged for 1st coauthorship on a paper that will be submitted, according to "International Publisher" LLC, to the special issue of a reputable journal of Frontiers Media. The value of coauthorship slots offered by "International Publisher" LLC with publication dates in 2019-2021 is estimated at $2.6 million. The price of all approximately 2000 papers offered over nearly three years reaches $6.5 million.

Why "International Publisher" LLC guarantees the publication of a manuscript despite the usual uncertainty of acceptance due to peer review can be explained by several reasons? First, according to the offers, in 2019 "International Publisher" LLC submitted manuscripts to journals with low impact factors (presumably predatory journals), where the probability of acceptance is high. Second, "International Publisher" LLC claims that it collaborates with journals and their editors and agrees upon the dates of publication. The offers can confirm these statements because some offers include the editor of the journal as a coauthor, which is mentioned in a comment on the offer. Third, according to the information on its website, "International Publisher" LLC also owns international journals, ensuring risk-free publication of the auctioned manuscripts (Litoy, 2019). Fourth, "International Publisher" LLC approaches legitimate authors to buy coauthorship in their high-quality manuscripts. It claims the following on its website:

> *"ко also work with foreign authors who publish the articles in good Q1-Q2 journals. The process looks like this: an author with a high Hirsch index writes an article to submit to a quality journal; one place is assigned to him; the remaining 2-3 places in the article are for sale. The payment is divided among the journal, the author, and us. Such schemes cannot be traced since there are only two sides, and each of them is interested in continuing cooperation".*

"International publisher" LLC also claims to sell coauthorship on manuscripts that are "already written and accepted by the journals". It is plausible in individual cases but I didn't find confirmation of this statement in the majority of identified papers. The analysis of offers showed that "International publisher" LLC uses cost-effective strategy, and manuscripts are written only after some co-authorship slots were sold. There is also evidence of announcements on different job search websites that orders to write a manuscript are placed in Ukraine. This gives a hint that most likely there won't be a request to change authorship after the acceptance of the manuscript.

Like many other broker companies, "International Publisher" LLC in its contracts mimics legitimate services providing "publishing services", e.g., "scientific journal selection" and assistance in the "publication of research" in journals[3].

Because both "International Publisher" LLC and users who purchase coauthorship demonstrate unethical behaviour, they attempt to maintain confidentiality. The titles of the journal and coauthors are available to scholars only after payment.

There is also a special condition in the contract:

> *"Each Party undertakes to maintain complete confidentiality of financial, commercial and other information received from the other Party. Such information could be transferred to Third Party only under the written consent of the both Parties, as well as in cases provided by law".*

"International Publisher" LLC uses aggressive marketing to attract potential clients. In addition to the website where the company impudently offers to sell coauthorship slots, it uses other malicious strategies. According to the website information, the company has contracts with different universities, and it organizes seminars for university faculties on publication strategies in international journals[4]. There is evidence of aggressive mail spamming of offers. The more that the company expands abroad, especially into the markets of post-Soviet countries, the Middle East and China, the more that it founds local offices. More than 10% of published papers are associated with China, Saudi Arabia, and the United Arab Emirates.

The nonethical activity of "International Publisher" LLC has received attention in media and blogs at both the national and international levels (Clarivate 2019, Litoy 2019, Marcus 2019, Chawla 2020, Abalkina 2021). However, its activity and consequences have not yet been investigated by scholars or by academic officials in Russia. In December 2021 Retraction Watch published the report by Perron et al. (2021) on the activity of "International Publisher" LLC. It focused on the communication with authors, journals and publishers concerning problematic papers from the paper mill. This current study is an independent research that sheds the light on the long-standing activity of the Russia-based

[3] See the archived example of the contract in Russian and English: https://web.archive.org/web/20220223120141/http://123mi.ru/1/contract.php?n=14&m=1

[4] https://www.bio.msu.ru/news/view.php?ID=1457

paper mill “International Publisher” LLC and identifies a set of predictors of fraudulent papers.

**Data**

The data were obtained from two main sources. First, since 2019, I collected the offers of “International Publisher” LLC published on the 123mi.ru/1 website. Second, the titles of papers were also provided in the contracts. I found that many contracts, especially those concluded by foreign offices, were not offered via their websites. Nearly every offer on the website 123mi.ru/1 comprises the topic of the paper, number of coauthorships for sale, price of a coauthorship, data about the journal (indexation in international scientometric database, quartile, country or region, scientific area), deadline for submission and approximate date of publication. “International Publisher” LLC does not openly disclose the title of the journal; the title is only available after payment.

“International Publisher” LLC started to publish offers of coauthorship for sale in mid-December 2018. Since then, more than 2000 offers of papers with coauthorship for sale have been created, and approximately 1000 papers, according to our estimation, have the publication deadline before March 2022. The website claims that, as of March 16, 2022, 5961 coauthorship slots had been sold.

**Identification of the papers**

The offers contain details that can facilitate the recognition of the published papers. First, the unique topic (title) of the article can provide sufficient information to identify the paper because many of them were published with identical or very close titles. The final result can be confirmed by the year of publication, country of the journal, indexation in international databases and the number of coauthorship slots. Please see Figure 2 as an example of the identification of a paper.

A total of 1009 offers and their titles were examined to detect auctioned papers published in journals. Each title from an offer was manually searched in Google, Google Scholar or Scopus. Some of the titles were found in Russian, so they were translated with Google Translate before the search.

As of mid-March 2022, 434 papers that potentially originated from the paper mill were identified. These 434 papers refer to 419 offers because one paper was published twice in different journals and 14 papers represent republication of a paper in hijacked journals without an offer with a separate number. The list of these papers can be accessed via https://docs.google.com/spreadsheets/d/1vzjtRPX7kd2KczdtKONEpRZb2F-

4lj5Sd9jL6DfbiBk/edit?usp=sharing. This list will be updated as soon as new papers potentially originating from this paper mill are identified. I included on the list only those problematic papers for which I have sound evidence of suspicious origin. The list does not include dozens of cases that were submitted with the intermediation of ”International Publisher” LLC but that were not offered through their 123mi.ru/1 website[5] or papers published in Russian journals. It also does not include problematic cases in which I doubted whether they came from “International Publisher” LLC.

[5] Except one title of the paper without an available offer. The title represented a republication of a paper published in a hijacked journal (see explanation below).

Figure 2

**Example of the identification of the offer**

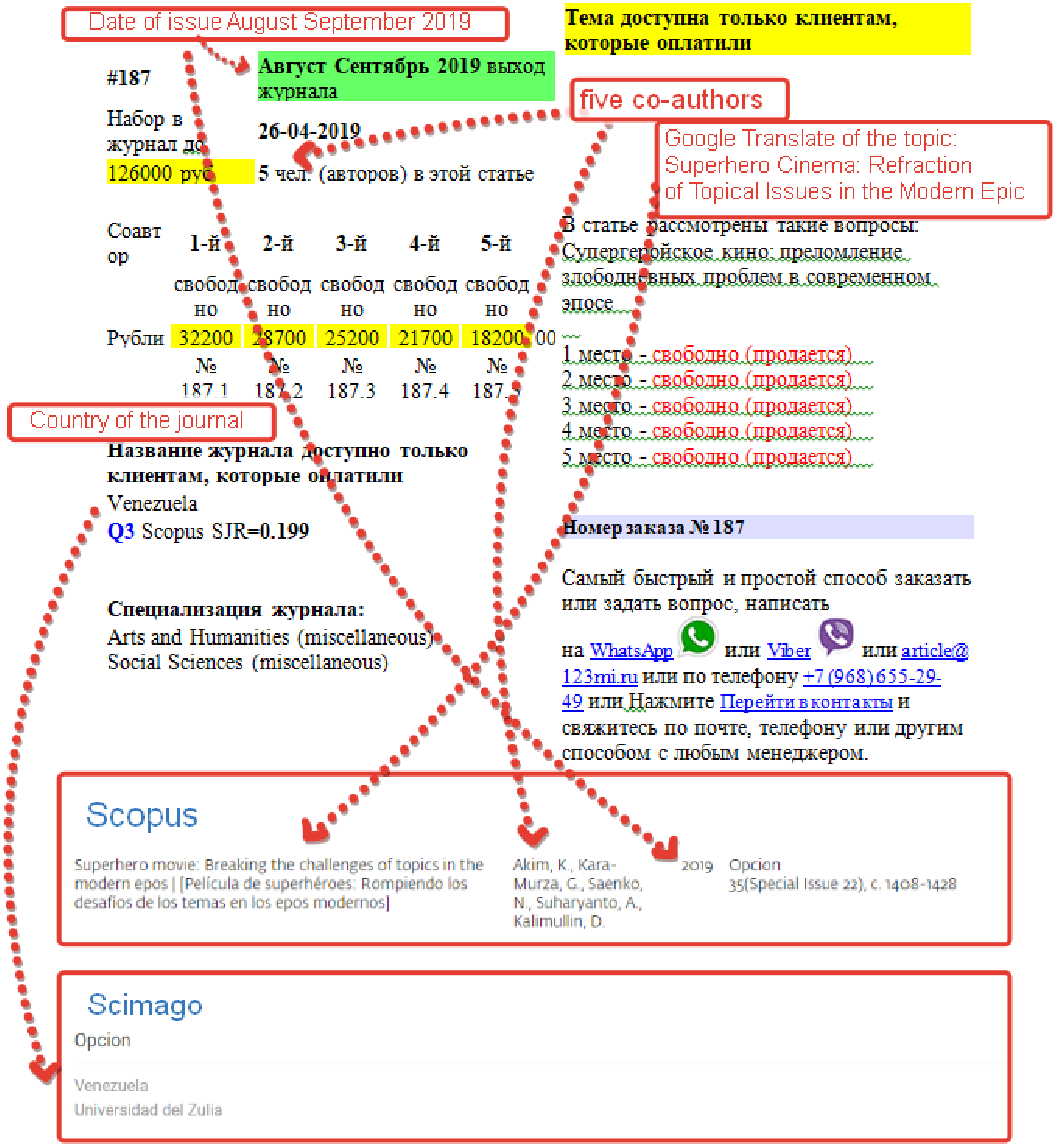

In some cases, it was a challenge to prove suspicious provenance. This was especially the case for papers with very common topics or more or less identical titles. I illustrate this difficulty with an identification and demonstration example (see Table 1).

**Table 1**

**Identification of papers**

| | **Offer** | **Title found** | **Title found** |
|---|---|---|---|
| Title | Supply chain and supply logistics as new areas of study in higher education | Supply chain and supply logistics as new area of study in higher education | Supply chain and logistics as new areas of study in higher education |
| Region/country of a journal | Europe | Germany | Venezuela |
| Number of coauthorship slots | 4 | 4 | 7 |
| Year of publication | 2020 | 2020 | 2020 |
| Decision | | Confirm | Reject |

*This is a demonstration. The titles are imaginary (the key words were changed to other terms) to avoid involving possible legitimate papers in this discussion.

I provide further evidence that these papers could be associated with the paper mill (see Figure 2).

For every paper, there is information about the deadline of publication and the number of coauthorship slots. The majority of offers on the website contain information about the region or country of the journal indexed in Scopus or Web of Science. Approximately 44% of identified papers have matches in these four parameters (title, equal year, equal number of coauthorship slots, equal country/region of a journal), 41% - three parameters and 15% - two. Approximately 47% of papers with three matches and less were additionally confirmed by coincidences of abstracts, and 33% - by country of coauthor which could provide sound evidence of questionable provenance of papers. There is only one paper on the list with only one match. I included it on the list of questionable papers because I have additional evidence of the affiliation country of the author. Moreover, this author is a coauthor of another paper on the list.

The mismatch of the number of coauthorship slots, date of publication or country of the journal can be explained in several ways. Fluctuations in the number of coauthors occur because not all coauthorship slots were sold or because it is possible to purchase an entire paper, e.g., all coauthorship slots, adding more coauthors to the paper or remaining a solo coauthor. Mismatches in the year of publication are rather rare (only 38 papers) and are mainly associated with earlier publication than was intended. In 35% of cases, the country/region of the journal does not correspond to the offer due to deindexation of the journal from Scopus or rejection of an article. I also suspect that, in some cases, “International Publisher” LLC was attracted by special offers from other journals and changed the target journal, as in the case of multiple submissions, to a hijacked journal instead.

## Collaboration anomalies

### 1. Suspicious collaborations

Under normal conditions, collaboration occurs on the basis of mutual scientific interests and research on similar topics and leads to joint work. Such collaborations are not accidental and involve personal acquaintances. In the case of coauthorship for sale, the purchase of slots of the manuscript occurs independently. A scholar pays for his or her coauthorship slot and order in the paper without knowing the other coauthors. Thus, the purchase of slots for the same paper occurs independently by authors who, in many cases, do not know each other. This phenomenon was also mentioned in the report by RAS (2020).

The patterns of such collaboration can be observed in the variety of affiliations among the authors of the article. Moreover, in many cases, the “authors” of an individual manuscript specialize in different disciplines that also do not correspond to each other and/or to the topic of a paper. In other words, the phenomenon of suspicious collaboration supposes a collaboration of scholars who: 1) might not be familiar with each other; 2) do not have common research interests; 3) are affiliated with different universities; 4) specialize in different disciplines; 5) and might not specialize in the topic of the paper. A striking example of such collaboration is a paper written by scholars from an economic University A and a medical University B on the topic of chemical engineering. Such a collaboration pattern is not misconduct itself but can serve as a predictor of violation of academic ethics and suspicious origin from a paper mill.

Suspicious collaborations can also be observed at the university level. Suspicious coauthorship includes collaborations in which the coauthors are affiliated with different organizations that might not engage in joint scientific cooperation. These cases can be detected by the comparison of collaboration in Russian and international journals. For example, according to the Russian scientometric database e-Library, the above mentioned medical University B is outside the top 100 collaboration organizations of the economic University A. In contrast, according to Scopus data, the medical University B reached the sixth rank among the top collaboration organizations of the economic University A. These data and such a mismatch suggest that such cooperation most likely represents artificial collaboration for the purpose of publication in international journals to inflate the publication record.

Another anomaly that can be observed in the collaborations linked to the paper mill is the presence of the first authorship associated with China. Approximately 14% of problematic papers from our list have Chinese scholars as the first coauthor. This finding is likely related to the system of financial rewards in China, where the first author receives everything (Liu & Chen 2018).

Such collaboration anomalies can be explained by country-specific patterns. First, the new system of requirements for publications and effective contracts introduced in Russian universities has required more research output. This policy led to the destruction of scientific collaboration and its replacement by groups interested in publication in international journals (Guba 2022). Second, such artificial collaborations appear to share the financial costs of publication. This phenomenon was also observed in Ukraine (Mryglod et al. 2021).

## 2. Coauthorship-specific patterns: number of coauthors

The increasing number of publications by Russian authors has been accompanied by a declining share of single-authored papers and an increasing number of coauthors and affiliations per article. Matveeva et al. (2021) examined the trend of collaboration patterns in publications with fewer than ten coauthors and demonstrated that the average number of affiliations per publication by 21 universities of the Russian University Excellence Initiative (Project 5-100) increased from 2.2 in 2012 to 2.6 in 2016.

In contrast, the average number of affiliations of problematic papers that potentially originated from paper mills is 3.2, while the average number of coauthors is 3.9 per article. These data do not suggest increased collaboration between authors but rather anomalies in scientific collaboration of the sample, which in all likelihood was the result of the acquisition of coauthorship by independent scholars.

Another significant aspect of authorship patterns is the share of single-authored papers. According to Web of Science data, in 1993-2019, the share of solo papers by Russian scholars reached 16% (Chankseliani et al. 2021). Of course, coauthorship patterns are highly dependent on the relevant discipline. The largest share of single-authored publications can be found in the humanities. According to the Russian Science Citation Index, among the 100 most successful authors in terms of the number of publications in relevant disciplines, the share of single-authored papers in the majority of social and human sciences exceeded 50%. In economics and psychology, it is more than 40%, and the smallest share is registered in astronomy, physics and chemistry because these disciplines are characterized by large teams and even mega-collaborations (Handbook on Scientometrics 2021). In our sample of suspicious papers, there are only eight single-authored papers (~2%). "International Publisher" LLC sells a single authorship only in Russian journals that are not subject to serious demand. According to the evidence of the offers of "International Publisher" LLC, single authorship in papers in international journals can be mostly explained by the lack of purchases of other coauthorship slots.

### 3. Alphabetical order

There are different norms regarding how to order coauthors of papers in different disciplines. Many disciplines apply contribution-based approaches or seniority rules (Fernandes & Cortez 2020). Some disciplines, such as economics, mathematics, and high energy physics, use mainly an alphabetical order in scientific publishing (Frandsen & Nicolaisen 2010, Waltman 2012, Weber 2018, Fernandes & Cortez 2020).

The majority of papers potentially originating from the Russian paper mill have from three to five coauthors. Out of 426 papers, 378, or 88.7% (I excluded the papers with one author), did not follow the rule of alphabetical order, including papers related to economics and business. The lack of alphabetical order is a consequence of the slot-order principle and could serve as a predictor of problematic papers together with other suspicious patterns.

Each individual paper might seem legitimate unless I analyse all of the sample and identify some anomalies that could predict fraudulent papers. To conclude the results section, I provide the peculiar features and predictors of the Russian paper mill.

- Suspicious collaborations
    - Diversity of affiliations per paper
    - Specialization of the universities not corresponding with each other (financial universities with medical universities if the subject of the paper is not the economics of health care, for example).
    - Specialization of the authors does not correspond to the title of the manuscript
    - Affiliations of the authors do not correspond to the topic of the manuscript
- Lack of single-authored papers
- Lack of alphabetical order

- The use of commercial email *is not a sign* of a paper mill, as in many Chinese paper mills (Seifert 2021). Many legitimate scholars in Russia use their personal email addresses for submissions.
- Similar structure of the papers. Normally, the traditional IMRAD structure is used, in which M is frequently entitled "Materials and Methods".
- The majority of clients of the paper mill are affiliated with universities but not with research institutes, which are numerous in Russia.
- The majority of suspicious papers are associated with Russia, Kazakhstan, China, Ukraine, and the United Arab Emirates or combination of these countries (see Appendix 1).

**Journals**

Journals are a key element in the system of publication of papers originating from paper mills. Articles supposedly originating from the "International Publisher" LLC paper mill have been published in hundreds of different journals. Initially, "International Publisher" LLC focused on publishing in a limited number of low-quality and predatory journals, such as Opcion and Espacios. Later, these journals were deindexed from Scopus. According to the website of "International Publisher" LLC, in 2020, the company changed its strategy and invited legitimate scholars for collaboration to sell ready texts or coauthorship slots. There is evidence that legitimate scholars receive such dishonest offers (Hyndman 2020). This policy change can be explained by the instability of publications in predatory journals, which can be quickly excluded from international scientometric databases. Moreover, dishonest papers in predatory journals can be identified. An investigation by the Commission for Counteracting the Falsification of Scientific Research of the Russian Academy of Sciences showed that a number of articles published in predatory journals appeared to be the result of collaboration with paper mills (RAS 2020). The RAS report identified 259 publications with translated plagiarism and problematic coauthorship (RAS 2020).

Indeed, since the fall 2020 and 2021, priority was reoriented toward legitimate journals of reputable publishers (Elsevier, Springer Nature, Emerald, Wiley, Taylor & Francis, etc.). In addition, there was a significant increase in the number of journals in which papers from the paper mill were published. I identified problematic papers in a total of 152 authentic journals. Obviously, many legitimate journals are not aware of the submission of articles from the paper mill. Individually tailored articles are submitted separately to one journal. According to the data obtained, 98 legitimate journals have published one article from the paper mill (Figure 3). These data should be interpreted with caution because I did not identify all of the offers.

Analysis of the offers allowed us to conclude that "International Publisher" LLC is very careful to submit numerous papers to legitimate journals, limiting submissions to one or several papers per year, making it impossible for an individual journal to detect a problematic paper because a single paper can appear absolutely legitimate.

However, apparently, a number of journals still turn out to be corrupt. According to "International Publisher"'s website, the company wanted to acquire a number of journals. In addition, dishonest cooperation with editors of some journals was not excluded.

**Figure 3**

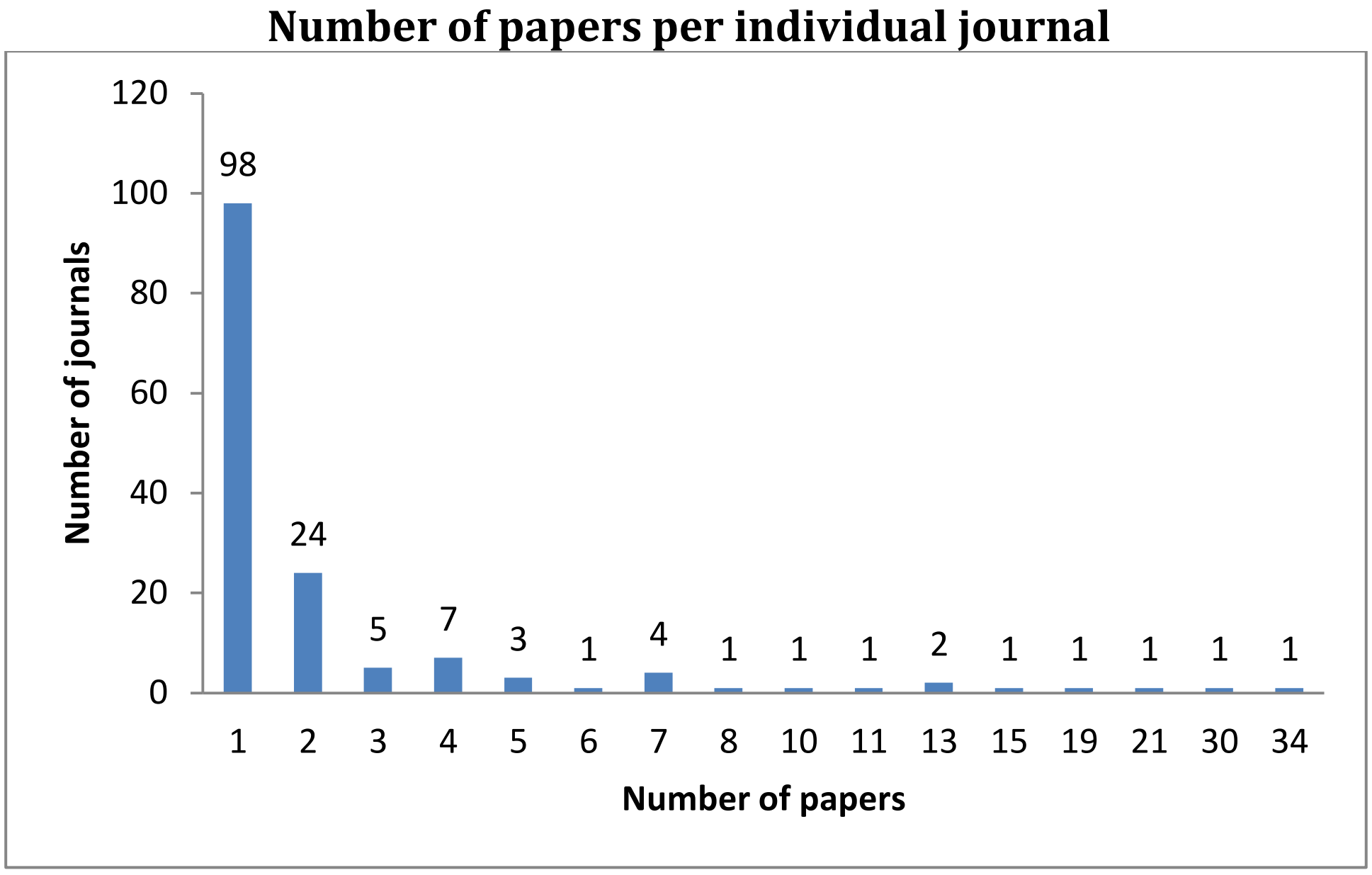

## 1. Questionable collaboration with journals

Appendix 2 shows that the greatest number of papers were published in one journal, the International Journal of Emerging Technologies in Learning. I found evidence of suspicious collaboration between the journal and the paper mill. In summer 2020, "International Publisher" LLC posted an offer:

> *"Special issue*
> *Note: this special issue will also contain papers #1081, #1082, #1083, #1084, #1085, #1086, #1087, #1088, #1089, #1090 on our website. A single author is free to purchase not more than 2 papers from this special issue tops. This is one of journals requirements."*

All ten papers were planned to be published in a German journal indexed in Scopus, Emerging Sources Citation Index, and EI Compendex. I detected nine of ten papers published in the same issue of the International Journal of Emerging Technologies in Learning, matching all of the journal characteristics in the offer (see Appendix 2). The same type of special issue offer was planned for the International Journal of Interactive Mobile Technologies, which belongs to the same publisher, International Association of Online Engineering, as the previous one. Appendix 3 shows five of 10 detected papers in the issue. The issue itself included only ten papers mostly "written" by Russian scholars that perfectly match the number of offers. However, I do not have sufficient supportive evidence to match the remaining five offers.

Such examples of two journals from the same publisher provide evidence of suspicious collaboration. It is highly unlikely that the journals are not aware of the questionable provenance of the papers. The share of identified problematic papers represents 3.0% of all papers indexed in Scopus by the International Journal of Emerging Technologies in Learning in 2020 and 2.3% in the International Journal of Interactive Mobile Technologies in 2020.

## 2. Questionable collaboration with editors

Analysis of offers and papers potentially originating from the paper mill allowed us to identify at least one episode of questionable collaboration between editors of MDPI journals and "International Publisher" LLC. Twenty of 21 identified papers published in MDPI journals had a specific feature: they were coauthored by scholars associated with one Eastern European country; Eighteen of them were affiliated with University C, and two had an affiliation with University D in this Eastern European country. One might suggest that these coauthors dishonestly purchased a coauthorship slot, but I suppose that the relationship is of a different nature. Some of these Eastern European coauthors were editors of several MDPI journals or guest editors of special issues. One could suggest that it is a coincidence, but some of the offers on the 123mi.ru website mentioned straightforwardly that one coauthorship slot of the paper was reserved for the editor of the journal or editor of the journal from this particular country. This coauthorship pattern in MDPI journals served as a good predictor of other dishonest papers.

Four MPDI journals (Sustainability (Switzerland), Journal of Theoretical and Applied Electronic Commerce Research, Energies, Mathematics) were involved in such suspicious collaboration patterns (see Figure 4).

**Figure 4**

**Suspicious collaboration patterns of MDPI journal editors**

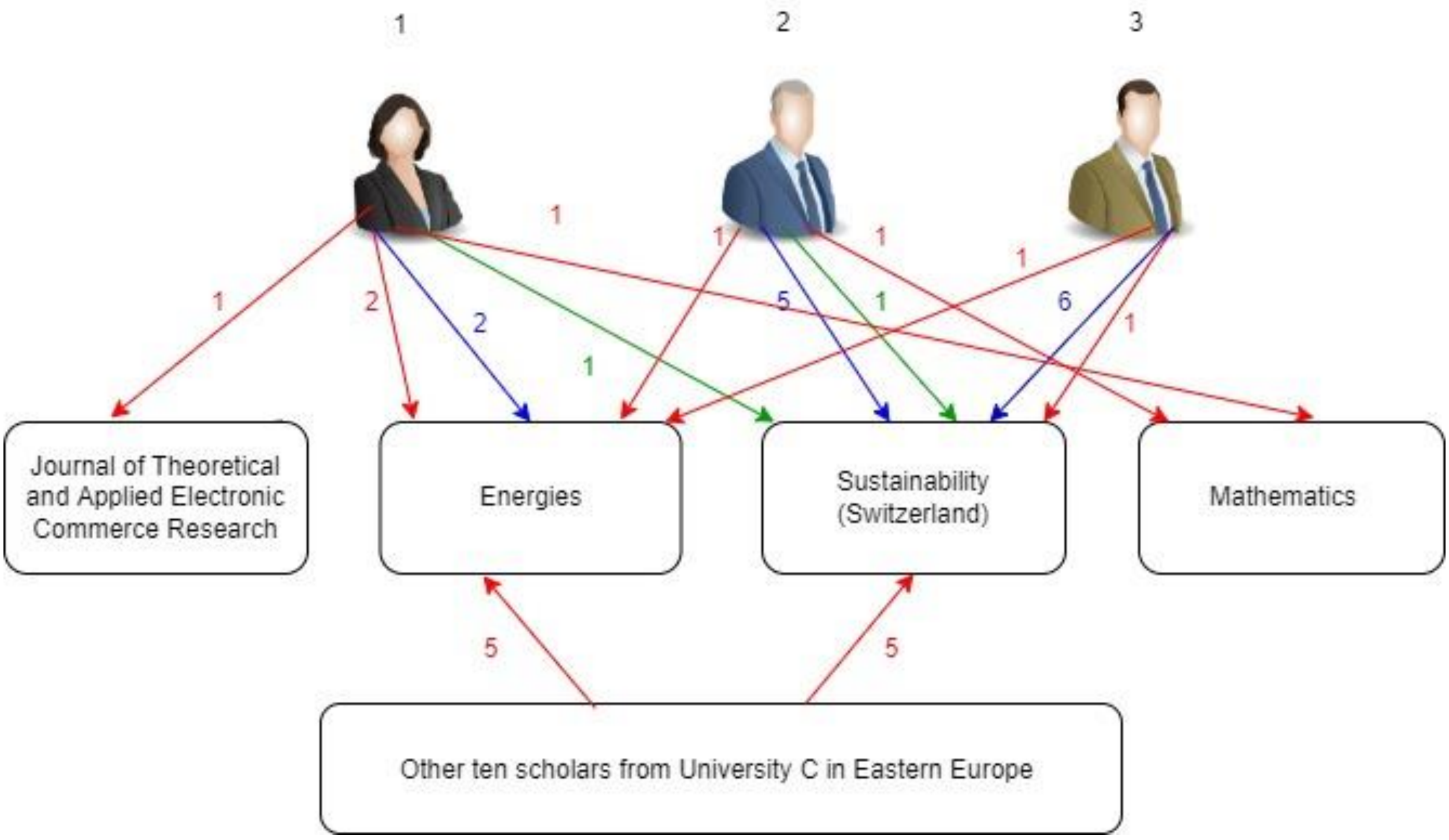

Red line - coauthorship in a paper potentially originating from the paper mill
Blue line - guest editor of a special issue
Green line – academic editor of a paper potentially originating from the paper mill.

At least eight papers potentially originating from the paper mill were coauthored by three editors of MDPI journals. All of them were affiliated with University C. Two editors were academic editors on one paper, likely purchased on the black market of academic papers. All three editors mentioned in Figure 4 were also guest editors of special issues of Energies and Sustainability (Switzerland), in which several papers of suspicious provenance were published. At least 10 other scholars from University C coauthored problematic papers that were published in MPDI journals.

## 3. Hijacked journals

Another pattern of papers potentially originating from the paper mill was detected. Twenty-two papers of questionable provenance were published in hijacked journals that mimic legitimate journals and fraudulently collect fees for rapid publication without providing peer review (Jalalian & Dadkhah 2015, Abalkina 2020, Moussa 2021). I detected such papers in three hijacked journals: Journal of Talent Development and Excellence (Abalkina 2020), Journal of Southwest Jiaotong University, and International Journal of Disaster Recovery and Business Continuity[6].

[6] All of the journals published at the website http://sersc.org/ are most likely hijacked journals.

Such collaboration between this broker company and fraudulent publishers provides further evidence for how hijacked journals work. Hijacked journals can publish thousands of papers over several months. Such numerous submissions are provided not only by aggressive marketing and spam emails but also by collaboration with national broker companies that have their own databases and clientele and accumulate papers for publication. Broker companies are attracted by the possibility of providing fast and guaranteed publication. However, broker companies themselves can be cheated by hijacked journals, as the case of “International Publisher” LLC shows. This conclusion is drawn from the detection of several republications of papers in hijacked journals with the same set of coauthors and on similar topics but with slightly different texts (see Table 2). A possible explanation for this republication might be that ”International publisher” LLC guaranteed the indexation of the published papers in Scopus or Web of Science, which is problematic in the case of hijacked journals. All nonlegitimate content from the Journal of Talent Development and Excellence was withdrawn by Scopus, and there is no evidence of indexation of papers from the Journal of Southwest Jiaotong University or International Journal of Disaster Recovery and Business Continuity. Therefore, contractual obligations forced to republish similar papers in legitimate journals, in most cases without advertising a respective offer on their websites. This republication of papers with the same set of coauthors and similar topics provides further evidence of possible provenance from the paper mill.

**Table 2**

**Republication of papers from hijacked journals**

| N | Title in a hijacked journal | Hijacked journal | Republished paper | Journal |
|---|---|---|---|---|
| 1 | Formation and Development of the Scientific and Scientific-Technical Activity Systems at Universities | Journal of Talent Development and Excellence, 2020 | Study of the system of scientific and scientific-technical activities of agrarian and economic universities*** | International Journal of Engineering Pedagogy, 2021 |
| 2 | Social Networks as a Means of Professional Communication | Journal of Talent Development and Excellence, 2020 | The Role of Social Networks in the Organization of the Educational Process and Learning*** | International Journal of Interactive Mobile Technologies, 2021 |
| 3 | Digital Economy as A Factor for Increasing the Competitiveness | Journal of Talent Development | Digital Development and Its Impact on Regions' Competitiveness*** | Global Economy Journal, 2021 |
| | of Countries and Regions | and Excellence, 2020 | | |
| 4 | Life Expectancy as an Economic Category: Social, Epidemiological and Macroeconomic Context | Journal of Talent Development and Excellence, 2020 | Population Aging and Its Impact on the Country's Economy* | Social Science Quarterly, 2021 |
| 5 | An Unmanned Aerial Vehicles Navigation System on the Basis of Pattern Recognition Applications | Journal of Southwest Jiaotong University/Xi'n an Jiaotong Daxue xuebao, 2020 | An unmanned aerial vehicles navigation system on the basis of pattern recognition applications—Review of implementation options and prospects for development*** | Journal of Software: Practice and Experience, 2021 |
| 6 | Internet Censorship in Developing Countries | Journal of Talent Development and Excellence, 2020 | Government regulation of the internet as instrument of digital protectionism in case of developing countries*** | Journal of Information Science, 2021 |
| 7 | Resource Sharing: Digital Economic Community Mediation | Journal of Talent Development and Excellence, 2020 | Formation and Implementation of a 'Digital Single Market' Concept in the Context of Digital Economy Expansion*** | Global Business Review, 2021 |
| 8 | Criteria for the Quality of Training of Future Specialists in Higher Educational Institutions | International Journal of Disaster Recovery and Business Continuity, 2020 | Determination of criteria for assessing the quality of training future specialists for higher education*** | International Journal of Educational Management, 2021 |
| 9 | The Impact of Information Technology on the GDP Growth Rate of Developing Countries | Journal of Talent Development and Excellence, 2020 | The role of information and communication technologies in a country's GDP: a comparative analysis between developed and developing economies** | Economic and Political Studies, 2021 |
| 10 | The role of the "Smart city" concept in the megacity municipal economy management | Journal of Talent Development and Excellence, 2020 | Municipal infrastructure management using smart city technologies*** | Theoretical and Empirical Researches in Urban Management, 2021 |
| 11 | Cloud Technology Development Alongside Public Life Digitalization | Journal of Talent Development and Excellence, 2020 | Assessment of the Impact of Cloud Technologies on Social Life in the Era of Digitalization*** | International Journal of Interactive Mobile Technologies, 2021 |
| 12 | E-learning for Promoting the Potential for Participating in Sports: Effective Physical Training for Children Aged 10 to 13 | Journal of Talent Development and Excellence, 2020 | Physical education and its influence on emotional and mental development of pre-schoolers*** | International Journal of Early Years Education, 2021 |
| 13 | International Student Exchange: The Practice of The BRICS Countries | Journal of Talent Development and Excellence, 2020 | Causes and Consequences of the Academic Migration from BRICS Countries to Developed Economies*** | Electronic Journal of Knowledge Management, 2021 |
| 14 | Native advertising in the media: a comparative analysis of ASEAN and BRICS | Journal of Talent Development and Excellence, 2020 | Going native: Prospects of native advertising development in the ASEAN and BRICS countries** | Mind and Society, 2022 |
| 15 | Digital Transformation of Social Institutions | Journal of Talent Development and Excellence, 2020 | Current trends in the digital transformation of higher education institutions in Russia*** | Education and Information Technologies, 2021 |

* - a separate offer was published on the 123mi.ru website.
** - an offer with the same offer number published on the 123mi.ru website.
*** - an additional offer was not found

## Discussion

The goal of our present study was to identify papers originating from the paper mill "International Publisher" LLC, analysing 1009 offers from the website 123mi.ru. I detected 434 papers that potentially originated from this paper mill. Unfortunately, I still did not recognize more than 500 other papers with forged authorship that infiltrated the academic literature.

Since I analysed only one paper mill, "International Publisher" LLC, there remains evidence of other paper mills in Russia and other post-Soviet countries (Marcus 2021). It is likely that the real number of paper mill production is much higher, and I detected only the tip of the iceberg. These fraudulent papers contaminate academic literature and they become more visible due to citations. Approximately the half of detected papers is cited at least once, and one problematic article was cited as much as 65 times as of March, 2022.

Such production of a Russia-based paper mill is difficult to detect due to individually tailored papers being submitted to more than one hundred different international journals. Journals themselves have no opportunity to notice irregularities from one single paper that can seem absolutely legitimate. This can require upgrading regularly the system of detection of fraudulent papers by publishers and by COPE. Current guidelines on detection of authorship for sale by COPE to trace similarity patterns between manuscripts or during the submission process are not helpful to identify individual fraudulent papers (COPE Council 2021). All the more suspicious collaboration should be taken in consideration. It is not misconduct itself but it can attract attention to the manuscript because there is evidence that papers originated from paper mills are accompanied by plagiarism and fabrication. This study sheds light on the patterns of the paper mill "International Publisher" LLC, which could help journals to identify suspicious papers.

The Russia-based paper mill went beyond the Russian scientific market having clients in at least 39 countries. More than 800 scholars affiliated with more than 300 universities were involved in the dishonest behavior connected with forged authorship. The majority of scholars are associated with just one coauthorship slot, though the leader coauthored 22 problematic papers. These numbers suggest that the increasing challenge of paper-mill activities and their proliferation across countries and universities.

The orientation towards publications in the journals indexed by Scopus and Web of Science has become a trap in the system of research output evaluation in Russia. The nationwide criteria require increasing publications from universities, and universities in turn motivate faculty to publish more to increase funding. Unfortunately, such a strategy, in addition to its advantages, transforms into a win-win strategy when faculty members with high workloads are unable to produce high-quality papers, but they can receive financial benefits with dishonest behaviour, while universities receive budget funding due to increased publication records. Many Russian and Chinese universities introduced financial rewards for publications. Though reverse situation was detected when researchers affiliated with Chinese or Russian institutions, having received a grant, buy co-authorship in a paper mill to demonstrate the output. Unfortunately, the difficulties to detect fraudulent papers and serious sanctions in some countries like Russia for violations of academic ethics only contribute to the proliferation of dishonesty.

## Conclusions

This study attempted to identify papers with forged coauthorship originating from the Russian paper mill "International Publisher" LLC by searching the paper titles from 1009 offers of coauthorship for sale and confirming the results by analysing the country of the journal, year of issue, and number of coauthorship slots. The major contributions of this paper are the following.

1) The current study allowed us to identify 434 suspicious papers that are most likely associated with the Russian paper mill “International Publisher”. Among these papers, there are at least fifteen republications of papers previously published in hijacked journals.

2) The Russian paper mill has a diversified strategy of collaboration with journals: a) one paper-one journal principle, e.g., submission of a problematic paper to an individual legitimate journal only once; b) submission to low-quality or predatory journals for which the rate of acceptance is rather high; c) dishonest collaboration with journals; and d) dishonest cooperation with the editors of journals.

3) The prevalence of dishonest papers from the Russia-based paper mill varies across the journals: the major share of papers from the paper mill are published in predatory journals or in journals with dishonest collaboration. More than 100 papers were published in journals of reputable publishers where mainly individual submissions were made. The evidence of the last year shows the reorientation of submissions to the journals of reputable publishers.

4) The activity of Russia-based paper mill has an international scope. The majority of papers potentially originating from the paper mill are mainly associated with Russia but also with Kazakhstan, China, Ukraine, and the United Arab Emirates. Scholars from at least 39 countries purchase coauthorship slots in Russia-based paper mill.

5) The analysis showed irregularities between the sample and common organization of science in Russia, providing further evidence of questionable provenance of the sample papers: a) suspicious collaboration between scholars affiliated with different organizations; b) topics of paper not corresponding to the specialization of the coauthors and their previous work; and c) the average number of coauthors in the sample being larger than it is typical in Russia, and vice versa, the number of solo papers being significantly smaller.

6) The present study provides further evidence of hijacked journal activity. This study demonstrates the strategies of hijacked journals in attracting potential authors through the intermediation of broker companies.

7) The current system of paper-mill detection should be regularly monitored and improved. This study provides predictors of a Russia-based paper mill.

**Acknowledgements**

I acknowledge the assistance of Sofia Ragozina in recognizing the papers; Evgeniy Enikeev for assistance in writing the script; and Andrey Zayakin for providing additional data. I am grateful to Guram Kvaratskhelia for technical assistance.

**Appendix 1**

## Number of purchased coauthorship slots by country (as of 16 March 2022)

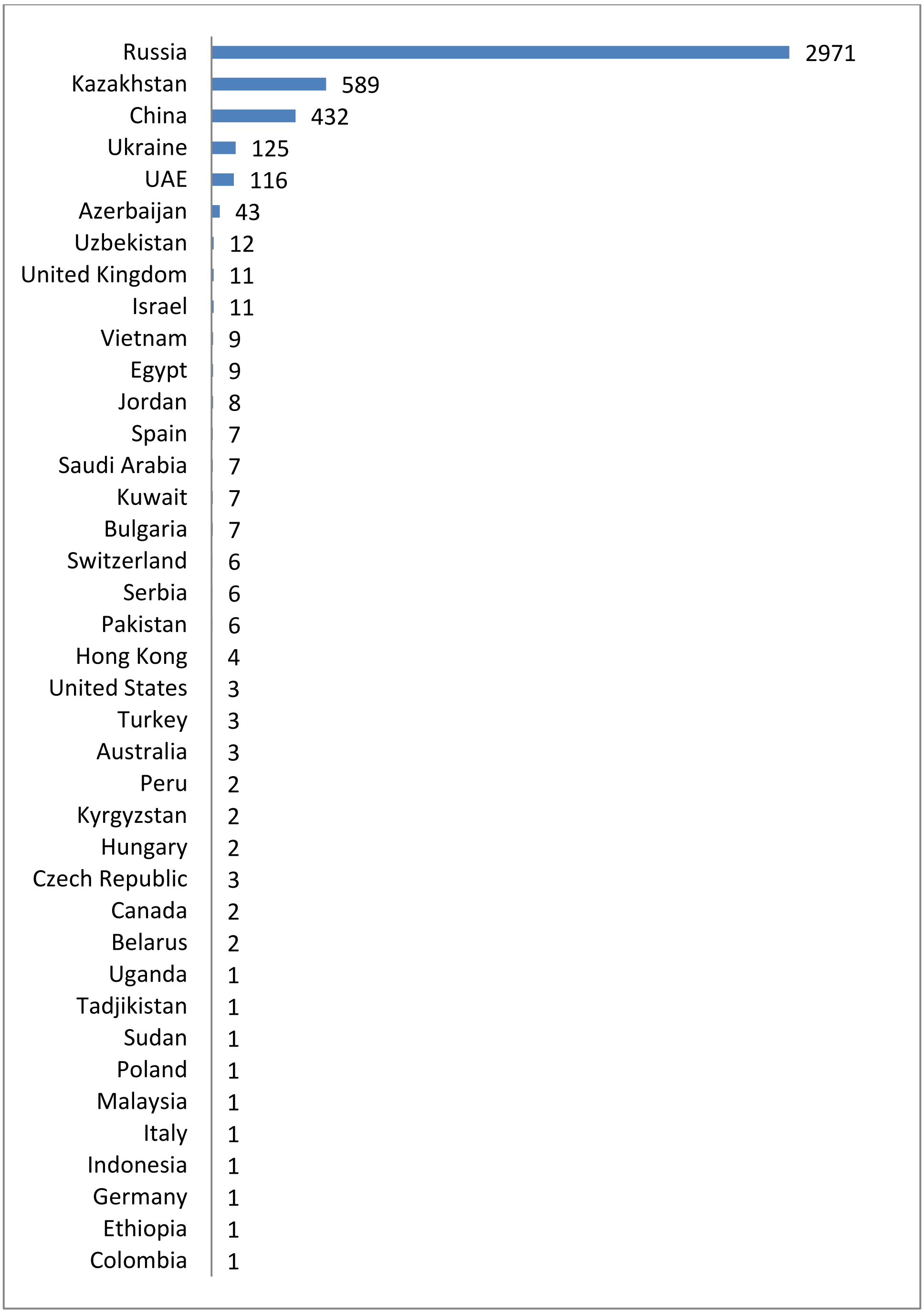

# Appendix 2
## An offer proposal for a special issue in a journal and identification of a special issue

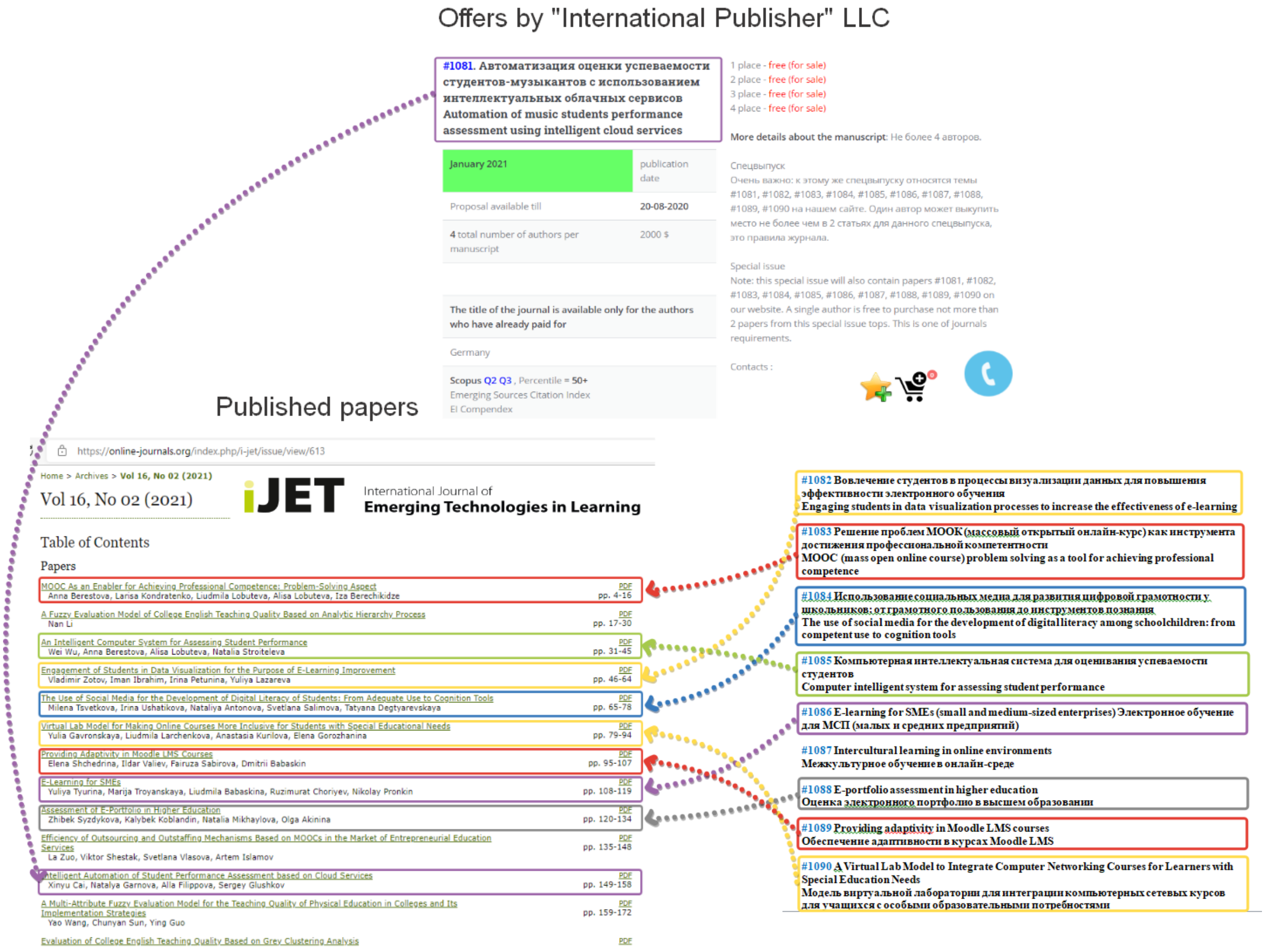

**Appendix 3**

## An offer proposal for a special issue in a journal

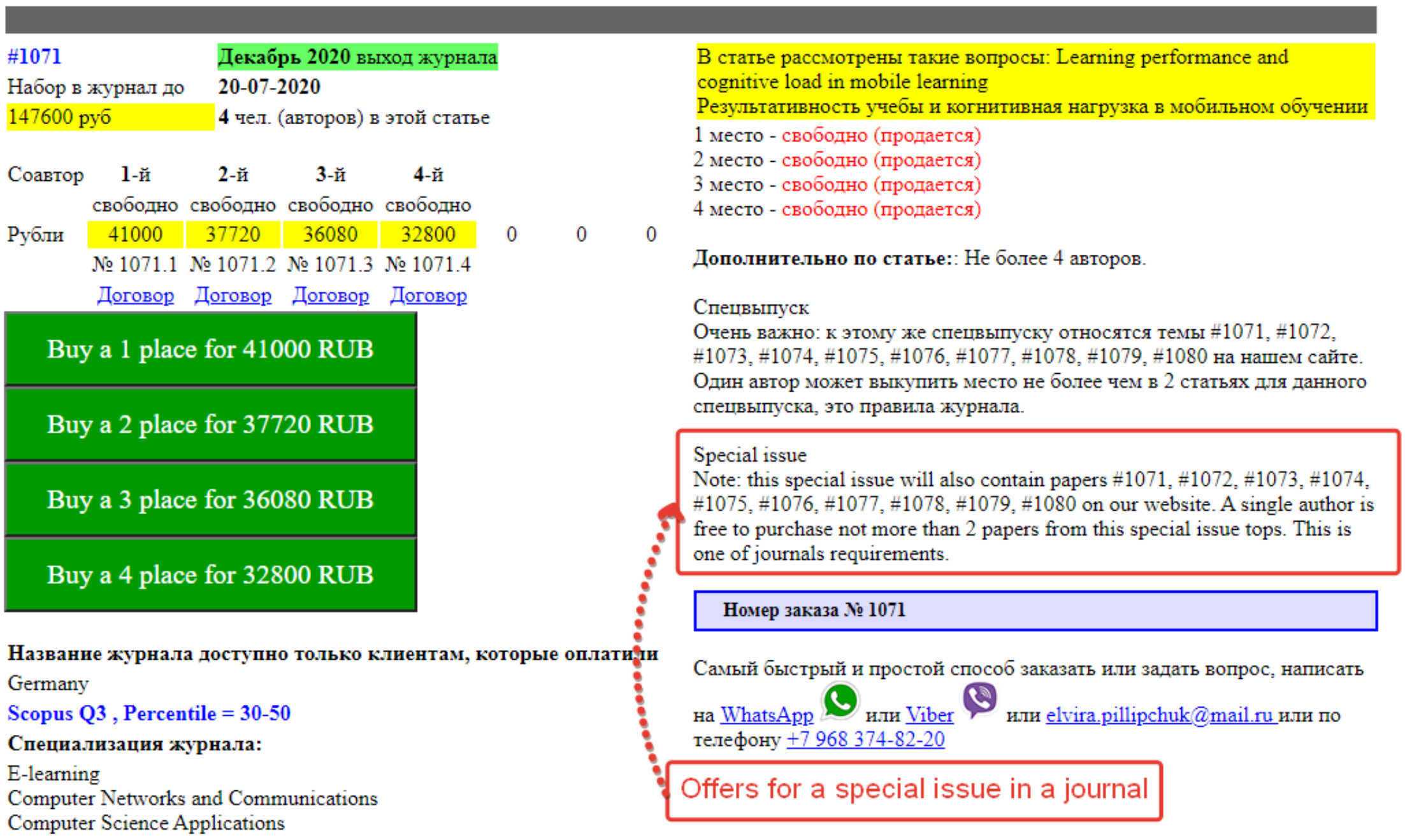

#1071 | Декабрь 2020 выход журнала
Набор в журнал до 20-07-2020
147600 руб | 4 чел. (авторов) в этой статье

| Соавтор | 1-й | 2-й | 3-й | 4-й | | | |
|---|---|---|---|---|---|---|---|
| | свободно | свободно | свободно | свободно | | | |
| Рубли | 41000 | 37720 | 36080 | 32800 | 0 | 0 | 0 |
| | № 1071.1 | № 1071.2 | № 1071.3 | № 1071.4 | | | |
| | Договор | Договор | Договор | Договор | | | |

Buy a 1 place for 41000 RUB
Buy a 2 place for 37720 RUB
Buy a 3 place for 36080 RUB
Buy a 4 place for 32800 RUB

**Название журнала доступно только клиентам, которые оплатили**
Germany
**Scopus Q3 , Percentile = 30-50**
**Специализация журнала:**
E-learning
Computer Networks and Communications
Computer Science Applications

В статье рассмотрены такие вопросы: Learning performance and cognitive load in mobile learning
Результативность учебы и когнитивная нагрузка в мобильном обучении
1 место - свободно (продается)
2 место - свободно (продается)
3 место - свободно (продается)
4 место - свободно (продается)

**Дополнительно по статье:**: Не более 4 авторов.

Спецвыпуск
Очень важно: к этому же спецвыпуску относятся темы #1071, #1072, #1073, #1074, #1075, #1076, #1077, #1078, #1079, #1080 на нашем сайте. Один автор может выкупить место не более чем в 2 статьях для данного спецвыпуска, это правила журнала.

Special issue
Note: this special issue will also contain papers #1071, #1072, #1073, #1074, #1075, #1076, #1077, #1078, #1079, #1080 on our website. A single author is free to purchase not more than 2 papers from this special issue tops. This is one of journals requirements.

**Номер заказа № 1071**

Самый быстрый и простой способ заказать или задать вопрос, написать на WhatsApp или Viber или elvira.pillipchuk@mail.ru или по телефону +7 968 374-82-20

Offers for a special issue in a journal

Source: 123mi.ru website

## Identification of papers in a special issue

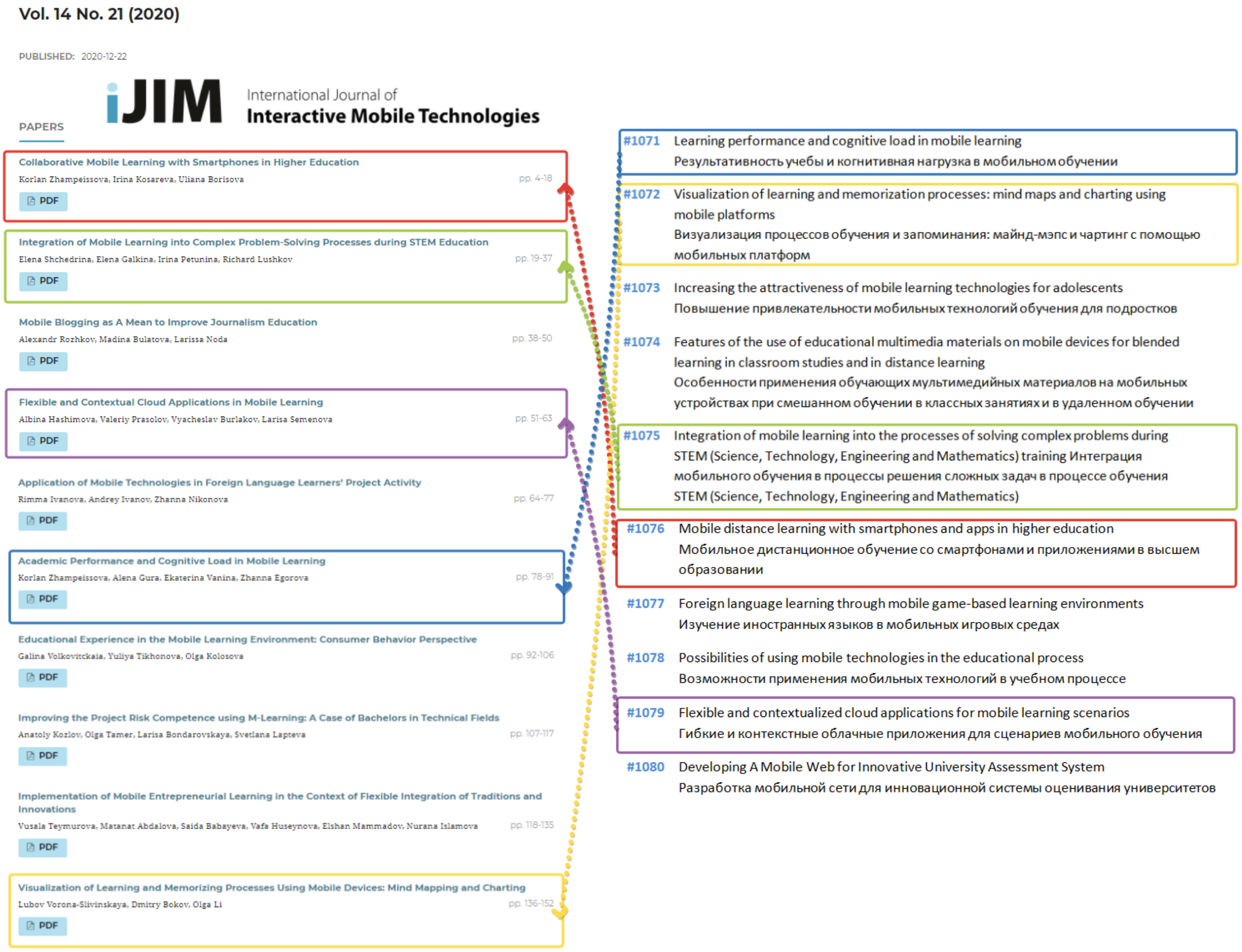

**Vol. 14 No. 21 (2020)**

PUBLISHED: 2020-12-22

iJIM International Journal of Interactive Mobile Technologies

PAPERS

Collaborative Mobile Learning with Smartphones in Higher Education
Korlan Zhampeissova, Irina Kosareva, Uliana Borisova — pp. 4-18 — PDF

Integration of Mobile Learning into Complex Problem-Solving Processes during STEM Education
Elena Shchedrina, Elena Galkina, Irina Petunina, Richard Lushkov — pp. 19-37 — PDF

Mobile Blogging as A Mean to Improve Journalism Education
Alexandr Rozhkov, Madina Bulatova, Larissa Noda — pp. 38-50 — PDF

Flexible and Contextual Cloud Applications in Mobile Learning
Albina Hashimova, Valeriy Prasolov, Vyacheslav Burlakov, Larisa Semenova — pp. 51-63 — PDF

Application of Mobile Technologies in Foreign Language Learners' Project Activity
Rimma Ivanova, Andrey Ivanov, Zhanna Nikonova — pp. 64-77 — PDF

Academic Performance and Cognitive Load in Mobile Learning
Korlan Zhampeissova, Alena Gura, Ekaterina Vanina, Zhanna Egorova — pp. 78-91 — PDF

Educational Experience in the Mobile Learning Environment: Consumer Behavior Perspective
Galina Volkovitckaia, Yuliya Tikhonova, Olga Kolosova — pp. 92-106 — PDF

Improving the Project Risk Competence using M-Learning: A Case of Bachelors in Technical Fields
Anatoly Kozlov, Olga Tamer, Larisa Bondarovskaya, Svetlana Lapteva — pp. 107-117 — PDF

Implementation of Mobile Entrepreneurial Learning in the Context of Flexible Integration of Traditions and Innovations
Vusala Teymurova, Matanat Abdalova, Saida Babayeva, Vafa Huseynova, Elshan Mammadov, Nurana Islamova — pp. 118-135 — PDF

Visualization of Learning and Memorizing Processes Using Mobile Devices: Mind Mapping and Charting
Lubov Vorona-Slivinskaya, Dmitry Bokov, Olga Li — pp. 136-152 — PDF

#1071 Learning performance and cognitive load in mobile learning
Результативность учебы и когнитивная нагрузка в мобильном обучении

#1072 Visualization of learning and memorization processes: mind maps and charting using mobile platforms
Визуализация процессов обучения и запоминания: майнд-мэпс и чартинг с помощью мобильных платформ

#1073 Increasing the attractiveness of mobile learning technologies for adolescents
Повышение привлекательности мобильных технологий обучения для подростков

#1074 Features of the use of educational multimedia materials on mobile devices for blended learning in classroom studies and in distance learning
Особенности применения обучающих мультимедийных материалов на мобильных устройствах при смешанном обучении в классных занятиях и в удаленном обучении

#1075 Integration of mobile learning into the processes of solving complex problems during STEM (Science, Technology, Engineering and Mathematics) training Интеграция мобильного обучения в процессы решения сложных задач в процессе обучения STEM (Science, Technology, Engineering and Mathematics)

#1076 Mobile distance learning with smartphones and apps in higher education
Мобильное дистанционное обучение со смартфонами и приложениями в высшем образовании

#1077 Foreign language learning through mobile game-based learning environments
Изучение иностранных языков в мобильных игровых средах

#1078 Possibilities of using mobile technologies in the educational process
Возможности применения мобильных технологий в учебном процессе

#1079 Flexible and contextualized cloud applications for mobile learning scenarios
Гибкие и контекстные облачные приложения для сценариев мобильного обучения

#1080 Developing A Mobile Web for Innovative University Assessment System
Разработка мобильной сети для инновационной системы оценивания университетов